\documentclass[prb,aps,epsfig,twocolumn,showpacs]{revtex4}

\usepackage{epsfig,amssymb,amsmath,latexsym}
\usepackage{subfigure}
\usepackage{wrapfig}
\usepackage{float}
\newcommand\beq{\begin{equation}}
\newcommand\eeq{\end{equation}}

\newcommand\bea{\begin{eqnarray}}
\newcommand\eea{\end{eqnarray}}

\newcommand\bi{\begin{itemize}}
\newcommand\ei{\end{itemize}}

\newcommand\non{\nonumber}

\newcommand\ie{{\it{i.e.}}}
\newcommand\etal{{\it{et al.}}}

\newcommand\ctd{{\textsf{CT}}}

\newcommand\ct{{\textsf{CT~}}}

\newcommand\db{{\textsf{DB~}}}

\newcommand\scbd{{\textsf{SB}}}

\newcommand\sdbd{{\textsf{SDB}}}

\newcommand\sdb{{\textsf{SDB~}}}

\newcommand\nsd{{\textsf{NS}}}

\newcommand\card{{\textsf{CAR}}}
\newcommand\ard{{\textsf{AR}}}

\newcommand\car{{\textsf{CAR~}}}
\newcommand\ar{{\textsf{AR~}}}

\newcommand\bdg{{\textsf{BdG~}}}

\newcommand\dc{{\textsf{DC~}}}

\newcommand\sar{{\textsf{SAR~}}}
\newcommand\sard{{\textsf{SAR}}}

\newcommand\scard{{\textsf{SCAR}}}

\newcommand\dbdg{{\textsf{DBDG~}}}
\newcommand\dbdgd{{\textsf{DBDG}}}

\newcommand\twod{{\textsf{2D~}}}

\newif\ifboo \boofalse

\begin{document}

\textheight=23.8cm

\title{Quantum charge pumping through a superconducting double barrier structure in graphene}

\author{Arijit Kundu$^{1}$, Sumathi Rao$^{2}$ and Arijit Saha$^{3}$}
\affiliation{
\mbox{$^1$ Institut f\"{u}r Theoretische Physik, Heinrich-Heine-Universit\"{a}t, D-40225 D\"{u}sseldorf, Germany}\\
\mbox{$^2$ Harish-Chandra Research Institute, Chhatnag Road, Jhusi, Allahabad 211019, India}\\
\mbox{$^3$ Department of Condensed Matter Physics, Weizmann Institute of Science, Rehovot 76100, Israel}\\
}

\date{\today}

\begin{abstract}
We consider the phenomenon of quantum charge pumping of electrons across a superconducting double barrier 
structure in graphene in the adiabatic limit. In this geometry, quantum charge pumping can be achieved by 
modulating the amplitudes ($\Delta_{1}$ and $\Delta_{2}$) of the gaps associated with the two superconducting strips. 
We show that the superconducting gaps give rise to a transmission resonance in the 
$\Delta_{1}$-$\Delta_{2}$ plane, resulting in a large value of pumped charge, when the pumping 
contour encloses the resonance. This is in sharp contrast to the case of charge pumping in a normal double 
barrier structure in graphene, where the pumped charge is very small, due to the phenomenon
of Klein tunneling. We analyse the behaviour of the pumped charge through the superconducting double barrier 
geometry as a function of the pumping strength and the phase difference between the two pumping parameters, 
for various angles of the incident electron.
\end{abstract} 

\maketitle

\section{\label{sec:one} Introduction}
The phenomenon of quantum charge pumping corresponds to a net flow of \dc current between different electron reservoirs
(at zero bias) connected via a quantum system whose parameters are periodically modulated in time~\cite{thouless,
bpt,brouwer}. The zero-bias current is obtained in response to the time variation of the parameters of the quantum 
system, which explicitly break time-reversal symmetry. It is necessary to break time-reversal symmetry in order to 
get net pumped charge, but it is not a sufficient condition. For obtaining a net pumped charge, parity or spatial 
symmetry must also be broken. Within a scattering approach, if the time period of modulation of the scattering 
system parameters is much larger than the time the particle spends inside the scattering region (dwell time), 
the adiabatic limit is reached. In this limit, the pumped charge in a unit cycle becomes independent 
of the pumping frequency. This is referred to as ``adiabatic charge pumping''~\cite{brouwer}. In  recent years,
quantum charge and spin pumping through various mesoscopic samples, involving quantum dots and quantum wires, 
have attracted increasing interest both theoretically
~\cite{pb,b3,b6,lev,ewaa1,saha,aleiner,sharma2,andrei,sela,das2005,amit,governale} and 
experimentally~\cite{marcus,leek,buitelaar1,giblin,blumenthal}, both in the adiabatic regime and otherwise.

The discovery of graphene, a two dimensional single layer of graphite, by K. S. Novoselov \etal~\cite{novoselovetal1} 
a few years ago, has led to an upsurge in the study of its transport properties, both theoretically and 
experimentally~\cite{geimreview,castronetoreview,sdsharmareview}. The low energy quasiparticle excitations in 
graphene behave like massless relativistic Dirac fermions. This provides us with an experimental test bed for 
observing many well-known phenomena in relativistic quantum mechanics, such as the Klein paradox~\cite{katsnelson} 
at low energies. In the recent past, a graphene-based quantum pump has been considered in literature
~\cite{RuiZhu,prada1,prada2} where pumped charge is obtained in an adiabatic quantum pump device based on a graphene 
monolayer modulated by two oscillating gate potentials. However, the pumped charge obtained in these kind of 
devices is quite small.

\begin{figure}
\begin{center}
\includegraphics[width=9.5cm,height=6.0cm]{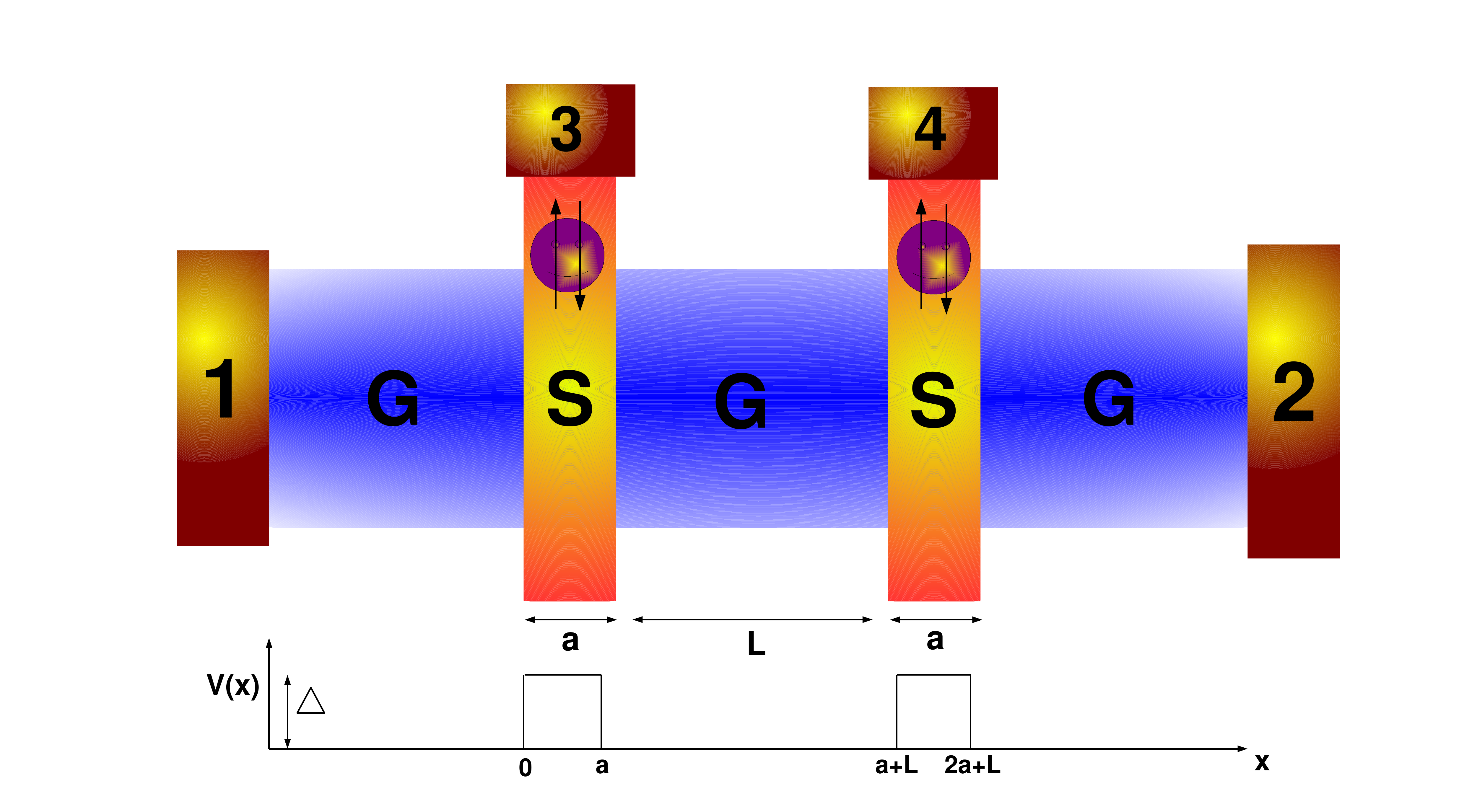}
\caption{(Color online) Cartoon of the \sdb structure where a  graphene
sheet is connected to two reservoirs  labeled 1 and 2. Superconducting material is
deposited on top of the two patches labeled S and 
connected to contacts labeled  3 and 4. The schematic of the 
potential profile seen by an incident electron is shown below.}
\label{figgeometry}
\end{center}
\end{figure}

Quantum charge pumps,  using a variety of  setups involving superconductors,
have also been  of major interest in recent years~\cite{wangsup,fzhou,blaauboer,governale1,taddei1,taddei2,kopvin,
wangwang,wangwang1,morpurgoPRL,sahadas}. Also, very recently, adiabatic charge pumping in graphene with superconductors
has also been considered~\cite{mblaauboer}. However an adiabatic quantum pump device based on superconducting double 
barrier (\sdbd) structures in graphene~\cite{kundusahaprb} has not yet been considered in the literature. 
Motivated by this fact, in this article we consider quantum pumping of electrons (in the adiabatic limit) across 
an \sdb structure in graphene, as depicted in Fig.~\ref{figgeometry}. Till date, no experiment has been carried 
out in the context of charge pumping with superconducting barriers. Experimentally it might be possible to design 
an \sdb structure by depositing thin strips of superconducting material on top of a single ballistic \twod graphene 
sheet at two places. This  can induce a finite superconducting gap in the barrier regions of the graphene sample 
due to the proximity of the superconducting strips. In principle, we can explore two scenarios to 
achieve significant amount of pumped charge $-$ {\textsl{(a)}} by periodic modulation of amplitudes 
$\Delta_1$ and $\Delta_2$ of the gaps at the two superconducting barriers (\scbd) or alternatively, 
{\textsl{(b)}} by periodic modulation of the order parameter phases $\phi_1$ and $\phi_2$ associated 
with the two barriers. In this paper, we explore the first alternative, since it has been seen
in earlier work~\cite{sahadas} that the second alternative leads to less pumped charge.
For free electrons in the graphene sheet, we show that in the $\Delta_1-\Delta_2$ plane, 
there is a sharp resonance point in the transmission probability across the \sdb structure. 
This is in sharp contrast to the case of charge pumping in a normal \db structure in graphene where the 
transmission probability across the \db structure does not have any resonance structure due to the phenomenon
of Klein tunneling. In the  \sdb geometry, when we consider $\Delta_1$ and $\Delta_2$ as the pumping
parameters, we can always choose a pumping contour which completely encloses the transmission resonance.
Hence it is possible for the  pumped charge to be large if the resonance is sharp enough. 

This paper is organized as follows. In Sec.~\ref{sec:two}, we discuss the theoretical modelling of the \sdb structure
in graphene and describe the resonance structure in the pumping parameter $\Delta_1-\Delta_2$ plane. In Sec.~\ref{sec:three}, we discuss the numerical results for the pumped
charge as a function of the various parameters in the theory and also contrast it with that of normal \db geometry. 
Finally in Sec.~\ref{sec:four}, we present our summary and conclusions.\\

\section{\label{sec:two} Resonant transmission in graphene \sdb structure}
Quantum transport in the \sdb structure in graphene was considered recently in
Ref.~\onlinecite{kundusahaprb}. Here we consider the same set-up, but instead
of applying a bias across terminals 1 and 2, we shall use terminals 3 and 4
to change the amplitudes of the superconducting barriers in a time-dependent
and out of phase fashion. The \sdb structure is formed by depositing thin strips of superconducting material 
on top of the graphene sheet at two places. This induces a finite superconducting 
gap ($\Delta_{i}e^{i\phi_{i}}$) in the barrier regions as a result of the proximity effect.
($i$ refers to the index of the strips). The geometry is shown in Fig.~\ref{figgeometry}.
The spatial dependence of the order parameter (which also acts as a scattering potential for 
the incident electron) can be expressed as
\bea
V(x) &=& \Delta e^{i\phi} \Theta(x)
\Theta(-x+a) + \Delta e^{i\phi} \non\\
&&  \Theta[x-(a+L)] \Theta[-x+(2a+L)]
\label{potential}
\eea
where $a$ is the width of the superconducting barrier and $L$ is the distance 
between the two barriers. Here we assume that the spatial variation of the potential steps is 
slow on the scale of the lattice spacing so that inter-valley scattering is suppressed. 
$\Theta$ is the Heaviside $\Theta$-function, and we have taken $\phi_1=\phi_2=\phi$, 
since we will not be looking at supercurrents (Josephson effect) in this work. 

Following Ref.~\onlinecite{kundusahaprb}, we use a four dimensional version of the 
Dirac-Boguliobov-de Gennes equation (\dbdgd)~\cite{beenakker1} for electrons and holes 
which is given by
\bea
\left({\begin{array}{cc} 
           \vec{k}.\vec{\sigma}-U & \Delta \\
\Delta^{\ast} & -(\vec{k}.\vec{\sigma}-U) \\
           \end{array}} \right)\left( {\begin{array}{c} 
           u \\
v \\
           \end{array}} \right)=\epsilon\left( {\begin{array}{c}
           u \\
v \\
           \end{array}} \right)
\label{dbdg}
\eea
where, $U=U({\bf r})+E_{F}$, and the energy $\epsilon$ is measured from the Fermi 
level of the superconductor. We assume that $U({\bf r})= 0$ in the normal graphene 
region and $U({\bf r}) = U_0$, a constant, independent of ${\bf r}$ in the proximity
induced superconducting region. Note that we have defined dimensionless variables 
\bea
x\Rightarrow \frac{xE_{F}}{\hbar v_{F}}, ~~~y\Rightarrow \frac{yE_{F}}{\hbar v_{F}}, 
~~~k_{y} \Rightarrow \frac{\hbar v_{F}k_{y}}{E_{F}}, 
\nonumber \\
\Delta \Rightarrow \frac{\Delta}{E_{F}}, ~~~\epsilon \Rightarrow \frac{\epsilon}{E_{F}}~~{\rm  and} 
~~~U \Rightarrow \frac{U}{E_{F}}
\eea
to replace the original ones. 

The solution of the \dbdg equations~\cite{beenakker1}, describing electrons and holes 
with incident energy $\epsilon$ inside the normal graphene regions ($\Delta_{(i)}=0$), 
can be written as
\bea
\Psi^{e\pm}=\frac{e^{ik_{y}y\pm ikx}}{\sqrt{\cos\alpha}}\left({\begin{array}{c}
                 e^{\mp i\alpha/2} \\
\pm e^{\pm i\alpha/2} \\ 
0 \\
0 \\
\end{array}} \right) 
\label{states1}
\eea
\bea
\Psi^{h\pm}=\frac{e^{ik_{y}y\pm ik^{\prime}x}}{\sqrt{\cos\alpha'}}\left({\begin{array}{c}
                 0 \\
0 \\ 
e^{\mp i\alpha^{\prime}/2} \\
\mp e^{\pm i\alpha^{\prime}/2} \\
\end{array}} \right)
\label{states2}
\eea
where $\alpha=\sin^{-1}[{k_{y}/(\epsilon+1)}]$, $\alpha^{\prime}=\sin^{-1}[{k_{y}/(\epsilon-1)}], 
k=\sqrt{\epsilon^{2}-k_{y}^{2}}$ and $k^{\prime}=\sqrt{\epsilon^{2}-k_{y}^{2}}$. $\alpha$ is the 
angle of incidence of the incoming electron (with wave-vector $(k,k_y)$) and $\alpha^{\prime}$ is 
the angle of reflection of the Andreev reflected hole (with wave-vector $(k^{\prime},k_y)$). 
For retro Andreev reflection (\ard), $\alpha^{\prime},k^{\prime}$ have opposite signs from 
$\alpha,k$ whereas for specular Andreev reflection (\sard), they have the same signs. 
The change from retro \ar ($\epsilon <1$) to \sar ($\epsilon>1$) occurs at $\epsilon =1$ 
(in our dimensionless units).

\begin{figure*}[ht]
  \centering
  \subfigure[$k_{y}=0.0$]{\includegraphics[width=0.3\textwidth]{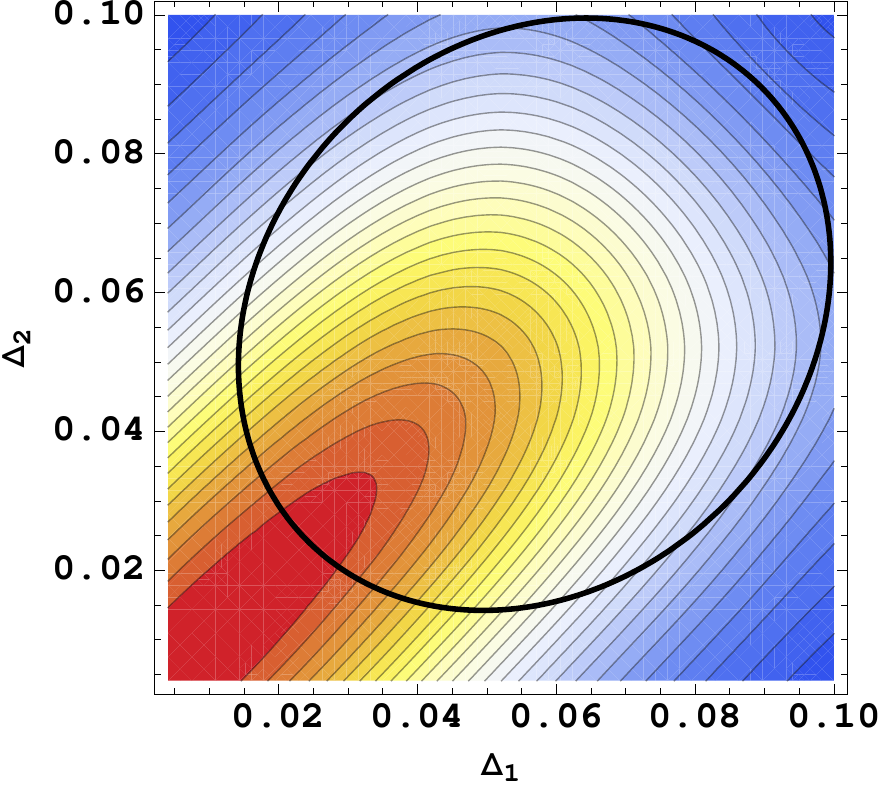}}
  \subfigure[$k_{y}=0.3$]{\includegraphics[width=0.3\textwidth]{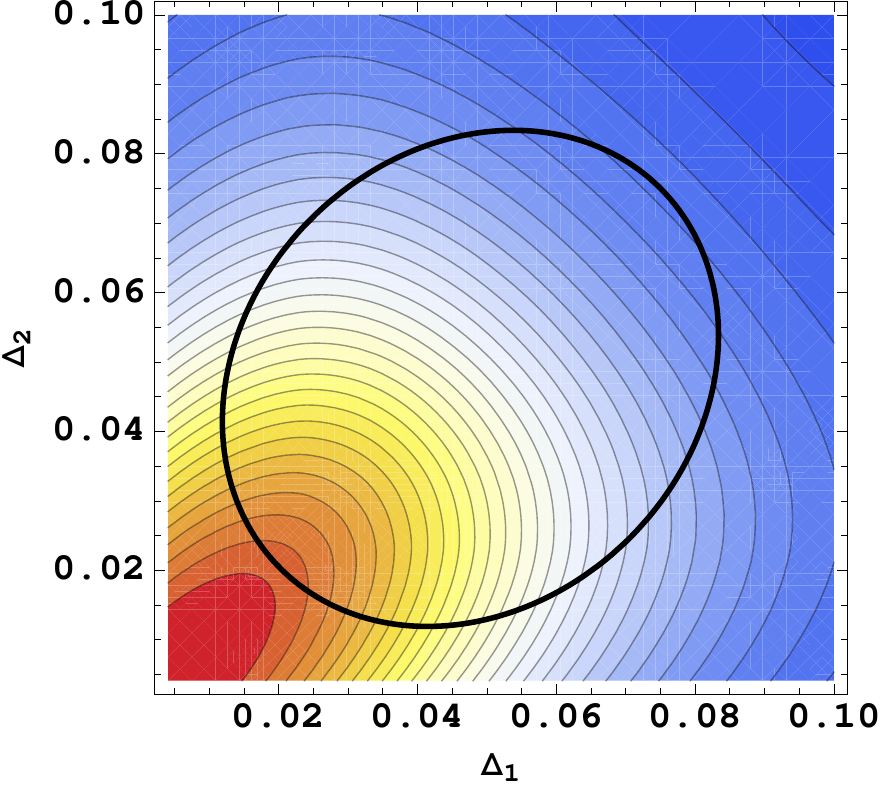}}
  \subfigure[$k_{y}=0.75$]{\includegraphics[width=0.3\textwidth]{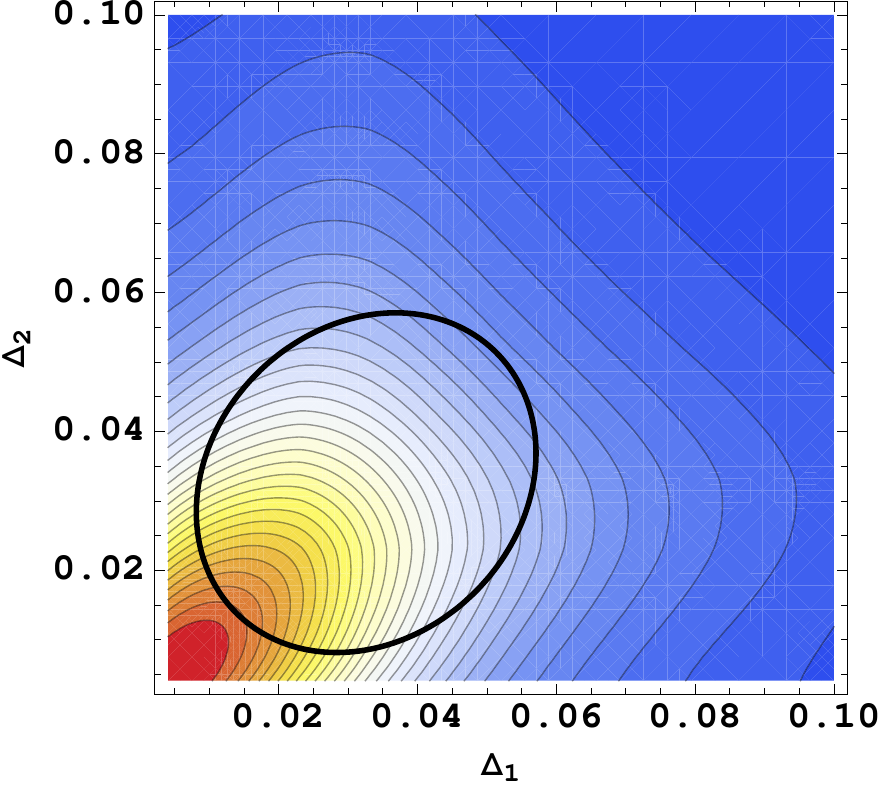}}
   \caption{(Color online) Contours of the transmission probability $|t_{c}|^{2}$ in the $\Delta_1-\Delta_2$ 
plane for three different values of $k_y$. 
The steps between the maxima and minima of $|t_{c}|^{2}$ range approximately 1.0 to
0.1, 0.6 to 0.05 and 0.3 to 0.01 for the three cases $k_y=0.0,0.3$ and 0.75 respectively.
In (a), (b) and (c)~$a/L=0.017, \Delta_{0}=40.0, 
U_{0}=10.0, \phi_{1}=\phi_{2}=0.0$ and $\epsilon=0.00142, 0.00121, 0.00082$ for the three contour plots
respectively. The black circle represents the pumping contour for the parameter values 
$\omega=1.0, P=30.0$ and $\eta=4\pi/9$.}
\label{figtwo}
\end{figure*}

Similarly for the superconducting barrier regions, the four component spinor solutions 
$(u,v)$ contain electron wave-functions $u$ of one valley and hole wave-functions $v$ 
of the other valley. The \dbdg equation can now be solved for any arbitrary energy 
$\epsilon$ and the four solutions inside the superconducting barriers are given 
in the preprint version of Ref.~\onlinecite{beenakker1} 
\bea
\psi_{1/2} &=& e^{ik_{y}y\pm x \sqrt{k_{y}^{2}-(U+ \sqrt{(\epsilon^{2}-\Delta^{2} )})^{2}}}\left( {\begin{array}{c}
                 e^{i\beta} \\
\pm e^{i\beta\pm i\gamma_{1}} \\ 
e^{-i\phi} \\
\pm e^{-i\phi\pm i\gamma_{1}} \\
\end{array}} \right) 
\label{wavefunction1/2}
\eea
\bea
\psi_{3/4} &=& e^{ik_{y}y \pm x \sqrt{k_{y}^{2}-(U- \sqrt{(\epsilon^{2}-\Delta^{2} )})^{2}}}\left( {\begin{array}{c}
                 e^{-i\beta} \\
\pm e^{-i\beta \pm i\gamma_{2}} \\ 
e^{-i\phi} \\
\pm e^{-i\phi \pm i\gamma_{2}} \\
\end{array}} \right) 
\label{wavefunction3/4}
\eea
where the subscripts $1/2$ refers to the upper and lower signs on the RHS respectively,
and similarly for $3/4$ 
and 
\bea
\gamma_{1}&=&\sin^{-1}\left(\frac{k_{y}}{U+ \sqrt{(\epsilon^{2}-\Delta^{2}})}\right) \nonumber \\
\gamma_{2}&=&\sin^{-1}\left(\frac{k_{y}}{U- \sqrt{(\epsilon^{2}-\Delta^{2}})}\right)
\eea
and 
\bea
\beta&=&\cos^{-1}\frac{\epsilon}{\Delta} \ \ \ if \ \ \epsilon<\Delta \nonumber \\
&=&-i\cosh^{-1}\frac{\epsilon}{\Delta} \ \ \ if \ \ \epsilon>\Delta~.
\eea
Here, we have not taken the limit $U \gg \Delta,\epsilon$.  We have also
obtained the solution for both right-moving and left-moving electrons and holes.

Solving the \bdg equation in the normal and superconducting regions, we obtain
the net quantum mechanical amplitudes for reflection, transmission (co-tunneling (\ctd)), 
\ar (and \sard) and crossed Andreev reflection (\card) (and specular crossed Andreev reflection 
(\scard)) of an electron incident on the \sdb structure, after it has traversed both the barriers. 

In the non-relativistic case,  a \db structure  always lead to resonances and this affects the 
transmissions and the reflections through the system. For relativistic electrons, 
the standard paradigm is that one cannot obtain confined carrier states for normal 
incidence~\cite{katsnelson,beenakker1} due to Klein tunneling. However, for relativistic
electrons with superconducting barriers, even for normal incidence, discrete Andreev bound levels 
can clearly lead to resonant transmissions in a \sdb structure in graphene~\cite{kundusahaprb}.

Here we use the standard wave-function matching technique to solve such scattering problems to 
obtain all the four quantum mechanical amplitudes across the \sdb geometry in graphene. 
To the left of the \sdb structure, with an incident electron from the left, the wave-function
can be written as
\beq
\psi^{e+}+r_{c}\psi^{e-}+r_{Ac}\psi^{h-}
\eeq
and to the right of the \sdb structure, the wave-function can be written as
\beq
 t_{c}\psi^{e+}+t_{Ac}\psi^{h+}~.
\eeq
Hence matching the wavefunctions in  the normal and proximity induced superconducting regions 
(Eq.(\ref{states1}-\ref{wavefunction3/4})) at the four normal$-$superconductor (\nsd) 
interfaces ($x=0,a,a+L,2a+L$) forming the \sdb structure, we obtain sixteen linear equations. 
Numerically solving these sixteen equations, we obtain the four amplitudes $r_c$, $r_{Ac}$, $t_c$ and $t_{Ac}$, 
for the \sdb structure, for an incident electron with energy $\epsilon$ below the gap $\Delta$~\cite{kundusahaprb}. 
Here $r_c$ is the reflection amplitude, $r_{Ac}$ is the \ar (and \sard) amplitude, $t_c$ is the 
\ct amplitude and $t_{Ac}$ is the \car (and \scard) amplitude. Note that we distinguish 
between electron and hole parameters, and hence the four amplitudes will be  different 
for incident electrons and holes. Moreover, in our numerical analysis we do not distinguish between 
specular and retro Andreev reflections.

\section{\label{sec:three} Numerical results for the pumped charge}

\begin{figure}[H]
  \centering
  \includegraphics[width=0.45\textwidth]{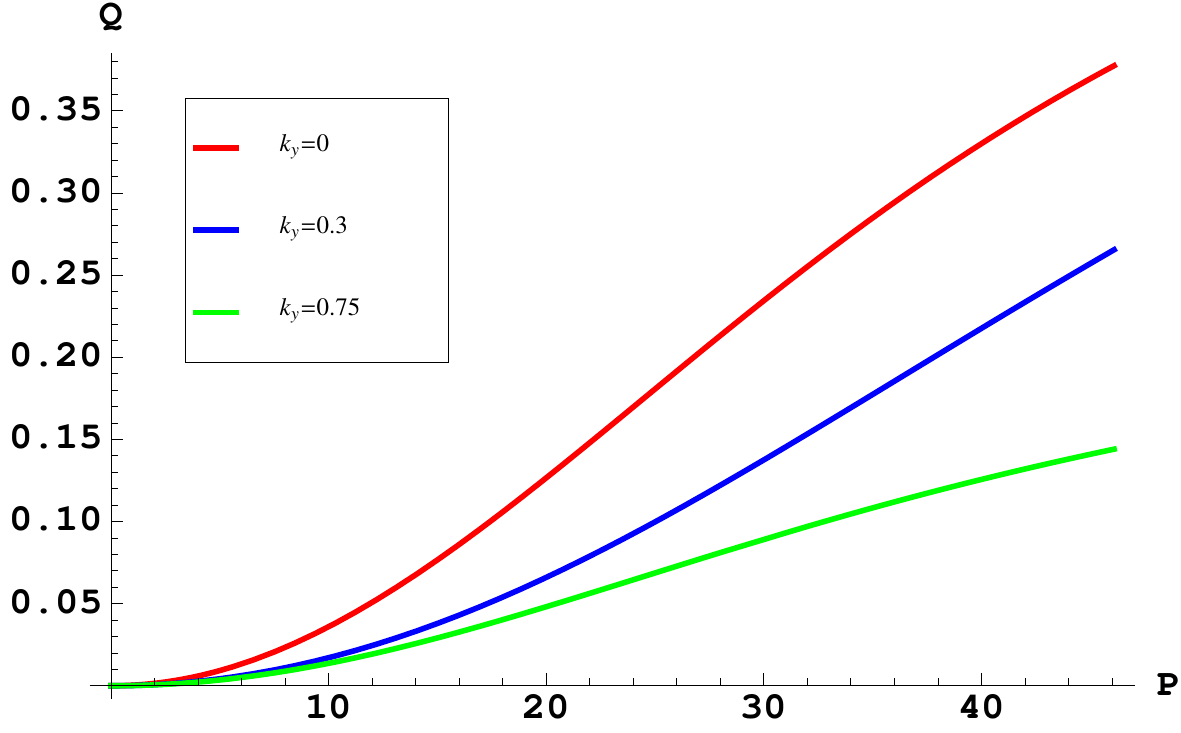}
   \caption{(Color online) The value of the pumped charge $\cal Q$ in units of electron charge $e$, for pumping in $\Delta_1-\Delta_2$ 
plane, is shown as a function of the pumping strength $P$ for three different values of $k_y$. Here $a/L=0.017, 
\omega=1.0, \eta=4\pi/9, \Delta_{0}=40.0$, $U_{0}=10.0$, $\phi_{1}=\phi_{2}=0.0$ and $\epsilon=0.00142, 0.00121, 0.00082$ 
for the three plots respectively.}
\label{figthree}
\end{figure}

In our numerical analysis for calculating the pumped charge through the \sdb structure, 
we choose the amplitudes of the order parameters of the two superconducting strips 
($\Delta_{1}$ and $\Delta_{2}$) as our pumping parameters. Here, $\Delta_{1}$ and
$\Delta_{2}$ are taken to oscillate with the frequency $\omega$, with a
modulation parameter $P$,  and a phase difference $2\eta$ between them, \ie, we choose
\bea
\Delta_{1} &=& [\Delta_{0} + P\cos(\omega t +\eta)]\epsilon \nonumber\\
\Delta_{2} &=& [\Delta_{0} + P\cos(\omega t - \eta)]\epsilon~.
\label{delta1delta2}
\eea
$\Delta_{0}\epsilon$ is the mean value of the amplitude around which the two pumping
parameters are modulated with time and $P$ is called the pumping strength. Here
$\epsilon$ is the energy of the incident electron. It is adjusted to be close to an Andreev bound
level, so that the \sdb is close to a resonance. Hence, effectively by varying $P$,
for fixed $\epsilon$ we vary the ratio $\epsilon/\Delta$. By fixing $\epsilon$ to be
close to the resonance, we maximise the pumped charge.
The presence of two time-varying potentials with a phase difference between them
explicitly violates parity, which is a necessary condition for obtaining pumped charge.
The frequency of the potential modulation is kept small in comparison  to the characteristic
times for traversal and reflection, so that the pump is in the adiabatic limit.

In Fig.~\ref{figtwo}(a), we plot the transmission probability ($|t_{c}|^{2}$) in the
$\Delta_1-\Delta_2$ plane for various values of the transverse electron momentum $k_{y}$. 
Unlike the case  of normal \db, where there is no resonance because
of the phenomenon of Klein tunneling, for the \sdb, $|t_{c}|^{2}=1$ resonance does occur for normal 
incidence of electrons, \ie~$k_{y}=0$. These bound states are produced due to the superposition of 
both electron and hole states and not just from any one of them. 
Hence these resonances are also different from the resonances that occur in normal double barriers
in other materials. For nonzero values of $k_{y}$, $|t_{c}|^{2}=1$ gets damped as normal reflection 
and \car (and \scard) also take part in transport with nonzero values~\cite{kundusahaprb} as shown 
in Figs.~\ref{figtwo}(b) and (c).

\begin{figure}[H]
  \centering
  \includegraphics[width=0.45\textwidth]{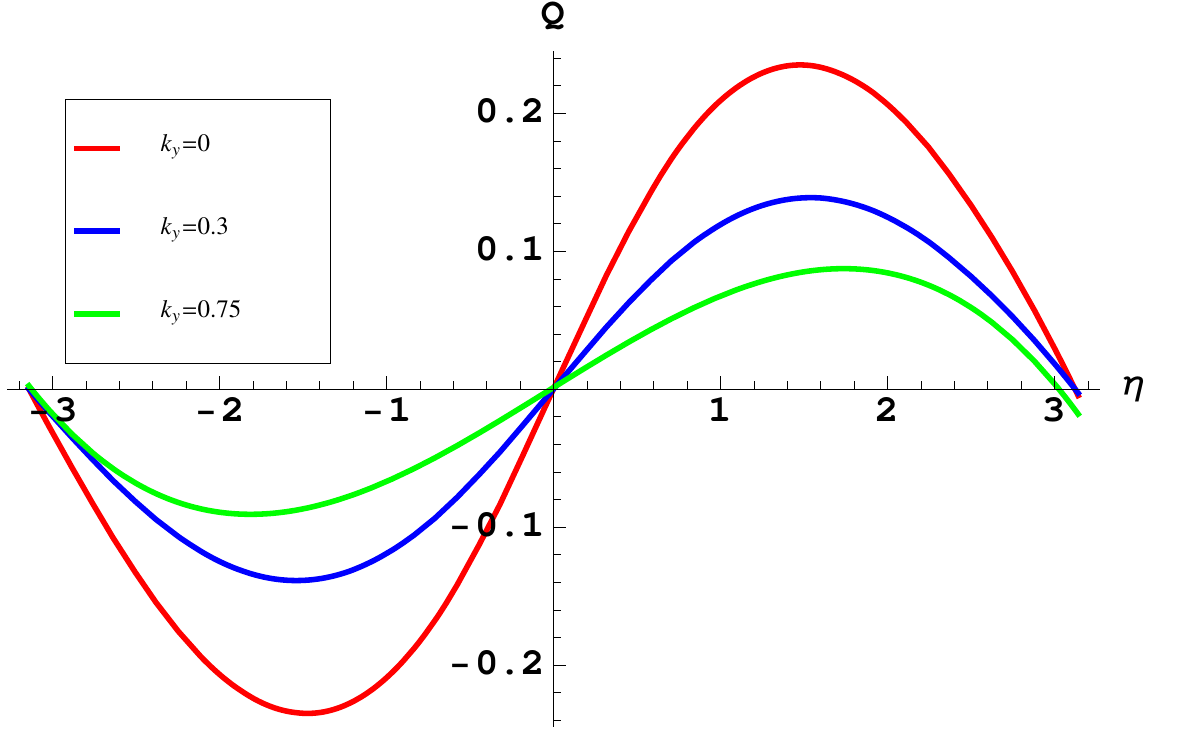}
  \caption{(Color online) The value of the pumped charge $\cal Q$ in units of electron charge $e$, 
for pumping in $\Delta_1-\Delta_2$ plane, is shown as a function of the phase difference between the two 
pumping parameters $\eta$ for three different values of $k_y$. Here $a/L=0.017, \omega=1.0, P=30.0, 
\Delta_{0}=40.0$, $U_{0}=10.0$, $\phi_{1}=\phi_{2}=0.0$ and $\epsilon=0.00142, 0.00121, 0.00082$ for the 
three plots respectively.}
\label{figfour}
\end{figure}

Using the modified version of the Brouwer's formula~\cite{brouwer,morpurgoPRL,sahadas}, the pumped 
charge through the graphene \sdb structure can directly be obtained from the parametric derivatives 
of the $\mathcal{S}$-matrix elements. For a single channel $\mathcal{S}$-matrix, we have
\bea {\cal Q} &=& {e\over 2 \pi} \int_0^\tau dt~
\Bigg[\Big(|r_{c}|^{2}\dot{\theta} + |t_{c}|^{2}\dot{\chi}\Big)\cos\alpha
\nonumber \\
&&~~~~~~~~~~\quad-\Big(|r_{Ac}|^{2}\dot{\beta} + |t_{Ac}|^{2}\dot{\gamma}\Big)\cos\alpha^{\prime}
\Bigg]
\label{pc} 
\eea
where $\theta$, $\chi$, $\beta$ and $\gamma$ correspond to the phases of the reflection, transmission,
\ar (and \sard) and \car (and \scard) amplitudes respectively. Note the negative sign in Eq.\ref{pc},
which results from the fact that $r_{Ac}$ and $t_{Ac}$ correspond to the conversion of an electron into a hole.
Thus, the pumped charge through the \sdb structure in graphene is directly related to the amplitudes 
and phases that appear in the $\mathcal{S}$-matrix. Inserting the unitarity relation $|r_{c}|^{2} + |t_{c}|^{2}
+ |r_{Ac}|^{2} + |t_{Ac}|^{2}=1$ in Eq.\ref{pc} we finally obtain
\bea {\cal Q} &=& {e\over 2 \pi} \int_0^\tau dt~
\Bigg[|r_{c}|^{2}\Big(\dot{\theta}\cos\alpha + \dot{\beta}\cos\alpha^{\prime}\Big)
\nonumber \\
&&~~~~~~~~~~\quad + |t_{c}|^{2}\Big(\dot{\chi}\cos\alpha + \dot{\beta}\cos\alpha^{\prime}\Big)
\nonumber \\
&&~~\quad + |t_{Ac}|^{2}\Big(\dot{\beta}-\dot{\gamma}\Big)\cos\alpha^{\prime}-\dot{\beta}\cos\alpha^{\prime}
\Bigg]
\label{pcf} 
\eea
Eq.\ref{pcf} is the working formula for pumped charge in our case. Note that if we substitute
$k_{y}=0$, (\ie~$\alpha=\alpha^{\prime}=0$) in Eq.\ref{pcf}, then the formula reduces to
\bea
{\cal Q} &=& {e\over 2 \pi} \int_0^\tau dt~\Bigg[|t_{c}|^{2} \Big(\dot{\chi} + \dot{\beta}\Big) - \dot{\beta} \Bigg]
\label{pcl}
\eea
which is precisely the modified Brouwer's formula used in Ref.\onlinecite{sahadas} for the quantum wire case.
In Eq.\ref{pcl}, the first term is called the ``{\it{dissipative part}}'' and the second term is known as the 
``{\it{topological part}}'' and depends entirely on the time derivative of the \ar phase.

\begin{figure}
\begin{center}
\includegraphics[width=8.5cm,height=6.0cm]{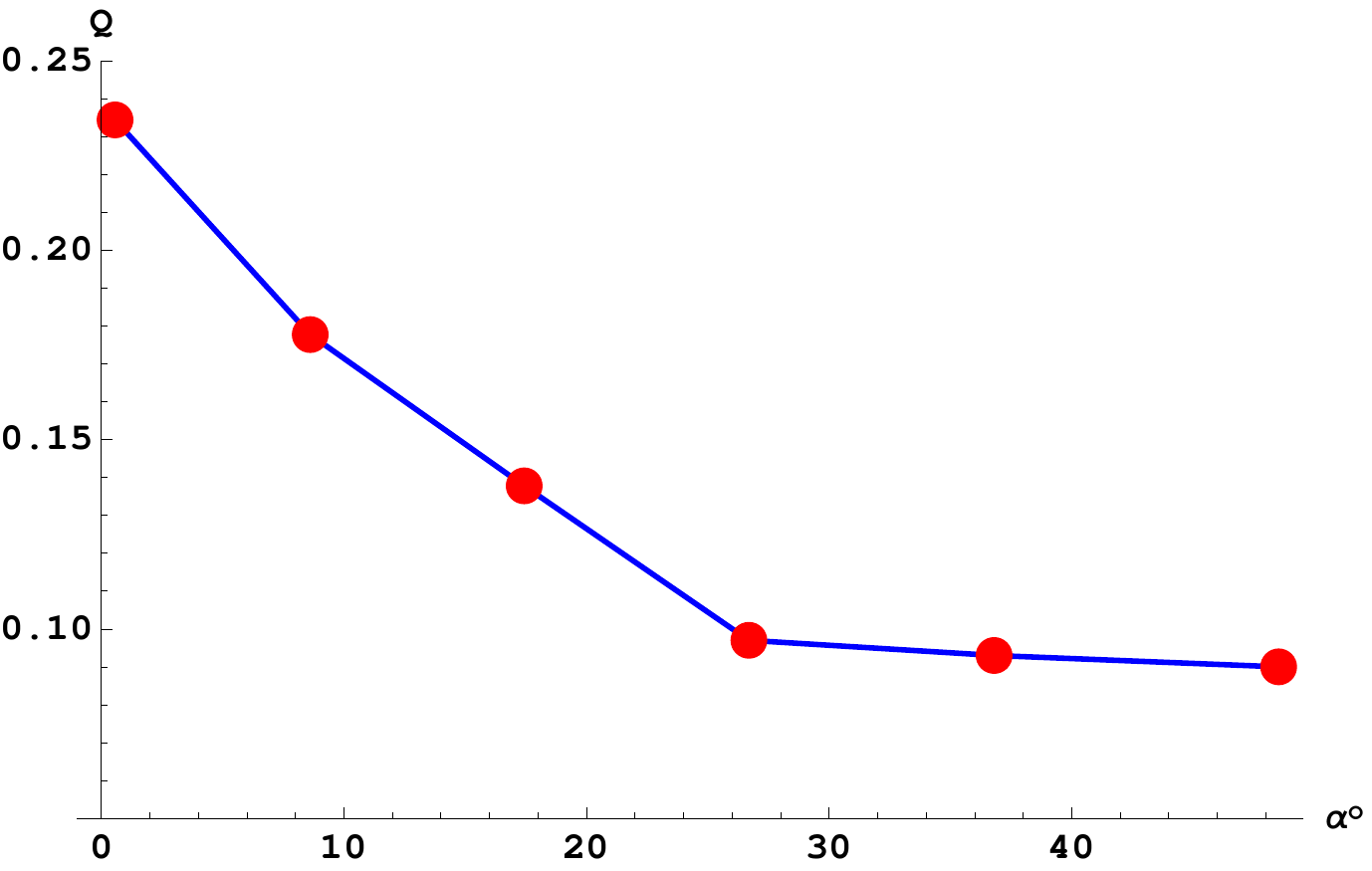}
\caption{(Color online) The maximum value of the 
pumped charge ${\cal {Q}}_{max}$ in units of electron charge $e$ 
through the \sdb structure, for pumping in $\Delta_1-\Delta_2$ plane, is shown as a function of 
the incident angle $\alpha$.}
\label{figfive}
\end{center}
\end{figure}

The pumped charge is obtained by using Eq.\ref{pcf} with  $\Delta_{1}$ and $\Delta_{2}$ as the two
pumping parameters. These two parameters are varied by periodically varying the top gate
voltage which controls the Fermi energy of the electrons in the superconducting region.
Thus essentially, as mentioned earlier, $\epsilon/\Delta$ is varied for the two barriers periodically. 
Using Eq.\ref{pcf}, we obtain the pumped charge for various parameters of the system. In Fig.~\ref{figthree} 
and Fig.~\ref{figfour}, we show the behaviour of the pumped charge as a function of the pumping strength $P$ 
and the phase difference between the two pumping parameters $\eta$ respectively, for three values of $k_{y}$. 
Note that in Fig.~\ref{figthree}(a), the pumped charge increases with the increase in the pumping strength 
$P$ as a larger value of $P$ corresponds to a larger pumping contour which encloses more and more of the resonance. 
Note also that  the $|t_{c}|^{2}=1$ resonance at $k_{y}=0$ is not a sharp resonance; it has a finite width because 
it also has contribution from the "{\it{dissipative part}}" as shown in Eq.\ref{pcl}. and is not purely topological.
The  "{\it {dissipative part}}" effectively reduces the pumped charge from integer values but we still obtain a fairly 
large value of pumped charge - where by large we mean that the pumped charge is a sufficiently large fraction of 
unity (roughly between $0.1$ and $0.35$). This is in sharp contrast to normal \db in graphene where the 
pumped charge~\cite{RuiZhu} is very small (in the range of $10^{-4}$, because there is no resonance due to Klein
tunneling. However for oblique incidence \ie~$k_{y}\neq 0$ (see Figs.~\ref{figthree}(b) and (c)), we obtain 
relatively smaller values of pumped charge as in this case normal reflection, \ctd, \ar (and \sard) and 
\car (\scard) also contribute to Eq.\ref{pcf} and the interplay between all the quantum mechanical amplitudes and 
their phases result in smaller value of pumped charge. 

In Fig.~\ref{figfour},  we show the oscillatory behaviour of pumped charge as a function of the phase difference 
$\eta$ between the two time varying parameters. Here also we note that we obtain smaller values of pumped charge 
through the \sdb geometry as we vary $k_{y}$ from normal incidence to oblique incidence. Note also 
that in Fig.~\ref{figfour}, for all three values of $k_{y}$, the pumped charge becomes maximum 
around $\eta \sim \pm \pi/2$~\cite{brouwer}. 

Finally, we also show the systematic behaviour of the maximum value of pumped charge through the \sdb geometry 
as a function of the incident angle $\alpha$ of the incident electron in Fig.~\ref{figfive}. This behaviour 
clearly shows that the maximum value of the pumped charge becomes smaller as we vary $\alpha$ from normal 
incidence ($\alpha=0$) to oblique incidence ($\alpha \neq 0$).

\section{\label{sec:four} Summary and Discussions}
To summarize, in this paper, we have studied adiabatic quantum pumping through
an \sdb structure in graphene and have shown that in the $\Delta_1-\Delta_2$ plane,  pumped charge
can be large (around 0.2- 0.3) in magnitude for normal incidence. This is in contrast to 
normal double barriers  in graphene, where the pumped charge is small (around $10^{-4}$)
due to the phenomenon of Klein tunneling. We also show that in the $\Delta_1-\Delta_2$ plane, 
transmission resonances ($|t_{c}|^{2}=1$) occur due to the formation of Andreev bound states between 
the two superconducting barriers. When the pumping contour encloses this $|t_{c}|^{2}=1$ resonance, 
we obtain significant amount of pumped charge for normal incidence ($k_{y}=0$). However, 
the transmission resonances get damped as the incident angle increases due to reflection and \sard. 
Hence, we obtain much smaller values of pumped charge for oblique incidence. We have also studied the 
pumped charge through the \sdb structure as a function of the pumping strength ($P$) and the phase difference
($\eta$) between the two pumping parameters. The most interesting study is the evolution of 
the maximum value of pumped charge through our proposed geometry as a function of the angle 
of incidence of the incoming electron. This shows monotonic decrease of pumped charge as we 
increase the angle of incidence of the incoming electron.

Similar behaviour has also been predicted for many other systems where one studies
quantum pumping through nanostructures. Integer pumped charge has been shown
for pumping through quantum dots~\cite{ewaa1,aleiner,saha} as well as through 
Luttinger liquids~\cite{sharma2,das2005,amit,sahadas}.
In more recent times, similar behaviour of pumped charge has been predicted in
graphene $NIS$ junctions~\cite{mblaauboer} and in an $InAs$ Josephson pump~\cite{fgiazotto}.

As far as the practical realization of such an \sdb structure in graphene is concerned, it 
should be possible to fabricate such a geometry by depositing thin strips of a spin singlet 
superconductors (like $Al$ or $Nb$) on top of a graphene sheet~\cite{hubert} at two places. 
The width of the strips should be of the order of the superconducting phase coherence length 
($10-15 nm$ in case of $Nb$) for \ct and \car (and \scard) to take place. In our 
geometry, pumped charge can be obtained by periodically varying the top gate voltage which 
controls the Fermi energy of the electrons in the superconducting region which  amounts 
to varying $\epsilon/\Delta$ for the two barriers periodically. The pumped current should
be in the range of pico-amperes when the pumping frequency is the order of a few MHz,
and should be experimentally measurable.

\acknowledgments{The work of A.S. was supported by the Feinberg Fellowship Programme at Weizmann, Israel.}

\bibliographystyle{apsrev} 

%
\bibliography{qpump_gsgsg_ref} 
\end{document}